\documentclass[a4paper]{article}

\usepackage[ninepoint]{itgspeech2025} %
\usepackage{times} %
\usepackage[english]{babel} %
\usepackage[utf8]{inputenc} %
\usepackage[T1]{fontenc} %
\usepackage[sort&compress,numbers]{natbib} %
\usepackage{microtype}
\usepackage{amsmath,amssymb,amsfonts}
\usepackage{array}
\usepackage{graphicx}
\usepackage[colorlinks=false,pdfborder={0 0 0}]{hyperref}
\usepackage[acronym,toc,shortcuts,nonumberlist]{glossaries}
\usepackage{xcolor}
\usepackage{xspace}
\usepackage{multirow}
\usepackage{subcaption}
\usepackage{hyperref}
\usepackage{cleveref}
\usepackage{siunitx}
\usepackage{pifont}
\usepackage[textsize=tiny,textwidth=1.4cm]{todonotes}

\newcommand*\nestedglsentry[1]{%
  \protect\ifglsused{#1}{%
    \glsentryshort{#1}%
  }{%
    \glsentrylong{#1}%
  }%
}
\newcommand{\authorrefmark}[1]{\textsuperscript{\footnotesize\ensuremath{#1}}}
\newcommand{\hours}[1]{\SI{#1}{\hour}}
\newcommand{\ms}[1]{\SI{#1}{\milli\second}}

\DeclareSIUnit\gigabyte{GB}
\newcommand{\gigabyte}[1]{\SI{#1}{\gigabyte}}

\makeatletter
\def\bstctlcite{\@ifnextchar[{\@bstctlcite}{\@bstctlcite[@auxout]}}
\def\@bstctlcite[#1]#2{\@bsphack
  \@for\@citeb:=#2\do{%
    \edef\@citeb{\expandafter\@firstofone\@citeb}%
    \if@filesw\immediate\write\csname #1\endcsname{\string\citation{\@citeb}}\fi}%
  \@esphack}
\makeatother

\newacronym{am}{AM}{acoustic model}
\newacronym{asr}{ASR}{automatic speech recognition}
\newacronym{ce}{CE}{cross entropy}
\newacronym{cnn}{CNN}{convolutional neural network}
\newacronym{ctc}{CTC}{connectionist temporal classification}
\newacronym{dnn}{DNN}{deep neural network}
\newacronym{fe}{FE}{feature extractor}
\newacronym{fir}{FIR}{finite impulse response}
\newacronym{fce}{f-CE}{frame-wise cross-entropy}
\newacronym{g2p}{G2P}{grapheme-to-phoneme conversion}
\newacronym{gelu}{GELU}{Gaussian error linear unit}
\newacronym{gpu}{GPU}{graphics processing unit}
\newacronym{gt}{GT}{Gammatone}
\newacronym{istft}{iSTFT}{inverse \nestedglsentry{stft}}
\newacronym{iir}{IIR}{infinite impulse response}
\newacronym{lm}{LM}{language model}
\newacronym{mfcc}{MFCC}{Mel-frequency cepstral coefficients}
\newacronym{nn}{NN}{neural network}
\newacronym{oov}{OOV}{out-of-vocabulary}
\newacronym{relu}{ReLU}{rectified linear unit}
\newacronym{sc}{SC}{supervised convolutional}
\newacronym{scf}{SCF}{supervised convolutional features}
\newacronym{stft}{STFT}{short time Fourier transform}
\newacronym{vgg}{VGG}{Visual Geometry Group}
\newacronym{wer}{WER}{word error rate}
\newacronym{werr}{WERR}{word error rate reduction}

\glsdisablehyper

\title{Unified Learnable 2D Convolutional Feature Extraction for ASR}
\author{Peter Vieting\authorrefmark{1}, Benedikt Hilmes\authorrefmark{1,2}, Ralf Schlüter\authorrefmark{1,2}, Hermann Ney\authorrefmark{1,2}}
\address{%
\authorrefmark{1}Machine Learning and Human Language Technology Group, RWTH Aachen University, Germany\\
\authorrefmark{2}AppTek GmbH, Germany\\
Email: \texttt{\{vieting,hilmes,schlueter,ney\}@hltpr.rwth-aachen.de}%
}

\begin{document}
\bstctlcite{IEEEexample:BSTcontrol}

\maketitle

\begin{abstract}
Neural front-ends represent a promising approach to feature extraction for \gls{asr} systems as they enable to learn specifically tailored features for different tasks.
Yet, many of the existing techniques remain heavily influenced by classical methods.
While this inductive bias may ease the system design, our work aims to develop a more generic front-end for feature extraction.
Furthermore, we seek to unify the front-end architecture contrasting with existing approaches that apply a composition of several layer topologies originating from different sources.
The experiments systematically show how to reduce the influence of existing techniques to achieve a generic front-end.
The resulting 2D convolutional front-end is parameter-efficient and suitable for a scenario with limited computational resources unlike large models pre-trained on unlabeled audio.
The results demonstrate that this generic unified approach is not only feasible but also matches the performance of existing supervised learnable feature extractors.
\end{abstract}

\noindent\textbf{Index Terms}: speech recognition, feature extraction, raw waveform modeling.
\glsresetall

\section{Introduction}
Current state-of-the-art \gls{asr} models use neural networks for acoustic modeling \cite{gulati2020conformer}.
In order for the \gls{am} to process speech, the waveform is transformed into an intermediate representation often called features.
This process traditionally consists of handcrafted static operations that take inspiration from psychoacoustic research on the human auditory system \cite{davis1980comparison,schlueter2007gammatone}.
Making the feature extraction a learnable part of the neural \gls{am} is desirable from a machine learning point of view, as it allows avoiding information loss by reducing the amount of heuristics and can be trained specifically for the needs of the subsequent neural model.
While successful approaches have been proposed \cite{sainath2015cldnn,sainath2015learning,zeghidour2018learning,tueske2018waveform,zeghidour2021leaf}, they are typically still inspired by traditional feature extraction methods.

This can manifest itself in a direct influence such as the initialization of parameters using handcrafted filterbanks \cite{zeghidour2018learning,sainath2015learning}.
Furthermore, parametric methods such as SincNet~\cite{ravanelli2018sincnet} rely on well-known and handcrafted signal processing techniques to constrain what the neural network can learn.
But even when training filters entirely from scratch, a more indirect influence such as the adoption of intuitions or insights from manual investigations may still be present.
One example is a network architecture design that resembles the structure of classical feature extractors, e.g. the Gammatone features in \cite{tueske2018waveform}.
This is also indicated by the observation that the filterbanks learned in this way resemble their classical counterparts \cite{tueske2014raw,tueske2018waveform,vieting2023itg}.
Furthermore, the usage of logarithmic or root activation functions \cite{sainath2015learning,zeghidour2018learning,tueske2018waveform} is in line with the non-linear human perception of loudness \cite{kollmeier2008perception} that is considered in classical features \cite{davis1980comparison,hermansky1990plp,schlueter2007gammatone} while it is not particularly common in neural networks elsewhere.
This raises the question, whether these influences are necessary and whether generic architectures are also feasible.

Many modern Conformer-based \gls{asr} models contain a convolutional subsampling block right after the feature extraction.
This is often implemented using a stack of VGG-style \cite{simonyan2015vgg} convolutions \cite{hori17_interspeech, zhang2017deepconv, watanabe2018espnet}.
Previous works on learnable front-ends replaced the classical feature extraction methods 1:1 in the architecture \cite{vieting2023itg}.
As a consequence, the typically 1D convolutional feature extractors are followed by the subsequent VGG-style 2D convolutional module.
From the perspective of architecture design, this results in a discrepancy between the different layer types, as the feature extractors and the VGG blocks do not blend in naturally.

There are two ways of unifying the currently used 1D- and 2D convolutional architectures.
The first is to remove the 2D layers and extend the generic 1D layers to also perform the subsampling to the final frame rate on which the \gls{am} operates.
This results in an architecture like the one used as the feature extractor in the wav2vec model family \cite{schneider2019wav2vec,baevski2020wav2vec2}.
\Cref{sec:learnable_feature_extractors} describes it in more detail.
Using the wav2vec feature extractor as a learnable front-end for standard supervised \gls{asr} has been studied in \cite{vieting2021waveform,vieting2023itg}.
These works refer to the usage of the convolutional feature extractor of wav2vec as a replacement for classical feature extraction, to keep the feature extraction lightweight compared to the \gls{am}.
The Transformer part, which makes up for most of the parameters of the wav2vec 2.0 model, is not considered.
This aligns with our goal to address a scenario with limited resources, where large pre-trained models such as the full wav2vec 2.0 or HuBERT \cite{hsu2021hubert} models used for representation learning are prohibitive.
Also, no additional unsupervised pre-training is used in this work.

We propose a second option for unification, namely the extension of the 2D convolutional layers to cover the full front-end.
In general, VGG-style models were successful by replacing convolutional layers with multiple layers with smaller kernels.
This is why we investigate whether this is feasible or even beneficial for learnable feature extraction in \gls{asr}.
\Cref{sec:2d} outlines the proposed concept.

Another aspect of generalization concerns SpecAugment \cite{park2019specaugment} that has become a de facto standard for regularization during training of \gls{asr} models.
As a further consequence of a 1:1 replacement of traditional features with learnable counterparts, SpecAugment is applied between the feature extractor and the VGG blocks in the remaining neural network \cite{vieting2023itg}.
From a general point of view, this is a rather arbitrary position inside the overall model.
Furthermore, it has several disadvantages for the learnable features \cite{vieting2025features}.
To avoid the usage of SpecAugment at this deliberate position inside the front-end, we can apply it in the \gls{stft}-domain before the feature extraction as proposed in \cite{vieting2025features}.
More details are provided in \Cref{sec:spec_aug}.

In summary, this work aims to reduce the influence of traditional feature extraction models on the design of the learnable front-end as much as possible.
Beyond final performance, we aim to unify and generalize the architecture of the neural front-end.
While similar work in \cite{vieting2023itg} trained learnable features without any access to spectral information for the first time, we take this yet a step further such that not even the structure is based on insights of previous manual investigations.
Not only does this mean that our studies combine both the feature extraction and the following VGG-style convolutions into a single unified front-end, but also SpecAugment is applied generically before the feature extraction instead of a rather arbitrary intermediate position.

The contributions of this work are as follows:
\begin{itemize}
    \item We propose a unified generic feature extraction architecture for \gls{asr} based on 2D convolutions that minimizes the influence of handcrafted methods,
    \item the results show that the proposed architecture is feasible and performs on par with existing learnable feature extractors and
    \item the analyses reveal that despite the generic design, the front-end shows behaviors that are consistent with long-established observations in the field.
\end{itemize}

\section{Feature Extraction Methods}
\vspace{-0.7ex}
This section presents the feature extraction methods used in this work.
As a classical, non-trainable baseline, we use \textbf{log Mel filterbank features}, which are arguably the most commonly used features for \gls{asr} nowadays.
They are computed by applying the \gls{stft} to the waveform, in this work with a window size of \ms{25} and a window shift of \ms{10}.
Subsequently, the squared magnitude is mapped into an 80-dimensional representation using the Mel filterbank.
The final features are generated by applying a logarithmic activation.

\vspace{-0.8ex}
\subsection{Existing Learnable Feature Extractors}
\vspace{-0.7ex}
\label{sec:learnable_feature_extractors}
All feature extractors are expected to be suitable for a low resource scenario.
Using large pre-trained models such as the full wav2vec 2.0 or HuBERT models is thus prohibitive in this context.
All learnable front-ends are jointly optimized with the entire \gls{am} using the supervised \gls{asr} training criterion on labeled data.
Here, we outline two existing approaches.

\textbf{Supervised Convolutional Features:}
The baseline learnable feature extraction in this work are the \gls{scf} as used in \cite{tueske2018waveform,vieting2021waveform,vieting2023itg, vieting2025features}.
The features are inspired by Gammatone features as the two convolutional layers resemble the Gammatone filterbank and the Hanning window used in \cite{schlueter2007gammatone}.
These layers are randomly initialized and serve as time-frequency-decomposition and temporal integration, respectively.
Unlike for Gammatone features, a multi-resolutional temporal integration is facilitated by the usage of multiple filters in the second layer.

The first layer operates on the waveform and consists of 150 filters with a size of \ms{16} and a stride of \ms{0.625}.
The activation function computes the absolute value.
The second layer consists of 5 filters with a size of 40 and a stride of 16, producing feature frames with a \ms{10} shift.
The final feature dimension is 750, as every filter is applied to the 150 output channels of the first layer.
The activation function is the 2.5\textsuperscript{th} root which is derived from the 10\textsuperscript{th} root used in Gammatone features and tuned for the learnable features \cite{tueske2018waveform}.
Layer normalization \cite{ba2016layernorm} is applied at the end.

\textbf{wav2vec Feature Extractor:}
The wav2vec framework \cite{schneider2019wav2vec,baevski2020wav2vec2} deploys an architecture with a convolutional feature extractor on the waveform.
While the general idea is the same in \cite{schneider2019wav2vec} and \cite{baevski2020wav2vec2}, the configuration differs slightly.
We follow the configuration from wav2vec 2.0 \cite{baevski2020wav2vec2}, which consists of seven 1D convolutional layers with 512 channels.
Additionally, group normalization is applied after the first layer and each layer is followed by a \gls{gelu} activation function \cite{hendrycks2018gelu}.
Since the Conformer operates on a frame shift of \ms{40} in our work, we add an eighth layer with the same configuration as the seventh layer to add another subsampling factor of two.
This results in layers with kernel sizes \{10, 3, 3, 3, 3, 2, 2, 2\} and strides \{5, 2, 2, 2, 2, 2, 2, 2\}.
Note that when referring to the wav2vec feature extractor, we mean the structure of the first convolutional layers as described above.
No other parts such as the quantization or Transformer modules are included here.
Also, no pre-training is conducted in this work and everything is trained from scratch with the supervised \gls{asr} criterion.

\vspace{-0.8ex}
\subsection{Generic Supervised 2D Convolutional Features}
\vspace{-0.7ex}
\label{sec:2d}
Our proposed 2D convolutional feature extraction is a generic front-end inspired by the commonly used VGG-style subsampling blocks in neural \glspl{am}.
Following their structural composition e.g. in \cite{zhou2022efficient}, it consists of a stack of layers that perform 2D convolution over the time and feature dimensions.
The additionally introduced channel dimension is merged into the feature dimension after the last 2D layer.
Similar to the VGG architecture or \cite{zhou2022efficient}, the kernel size is $3\times3$.
However, unlike \cite{zhou2022efficient}, no preceding classical feature extraction is used and we apply more convolutional blocks instead.
Exactly as many layers with strides of two in the time dimension are used so that the resulting output frame rate is \ms{40}.
More layers with a stride of one may be added in between.
For activation, we use the \gls{relu} function.

Special attention has to be devoted to the first layer.
Since the waveform is single dimensional, the first layer has to generate a feature dimension so that the subsequent layers can perform 2D convolution over time and feature dimensions.
This can be done via applying the \gls{stft}, using either the magnitude or real and imaginary parts.
In the latter case, the two parts are passed through a single 2D convolutional layer each and then summed before passing through the remaining 2D layers.

A second possibility is having a filterbank as a first layer.
It can be initialized similar to conventional feature extraction, e.g. with a Gammatone filterbank, and frozen during training.
However, to keep the pipeline as generic as possible, we also experiment with a learnable filterbank and random initialization.
To optimize the performance, we tune the kernel size and stride used by this filterbank in our experiments.

\vspace{-0.5ex}
\section{Experimental Setup}
\vspace{-0.5ex}
\subsection{Data}
\vspace{-0.5ex}
We conduct our experiments on the English dataset LibriSpeech \cite{panayotov2015librispeech}.
It consists of \hours{960} of audiobook recordings for training.
As labels, we use phonemes with the phoneme set consisting of ARPABET phoneme symbols without stress marker.
In order to generate phoneme sequences for words that are not contained in the given lexicon, we use Sequitur \cite{bisani2008g2p}.
During recognition, we use the official 4-gram \gls{lm} trained on the monolingual corpus data.
The final performance is evaluated on the standard dev and test sets provided with the dataset.

\vspace{-0.5ex}
\subsection{Training}
\vspace{-0.5ex}
A \gls{ctc} model is used to perform \gls{asr} in this work.
For models with log Mel and \gls{scf} features, the feature extraction is followed by a VGG-style convolutional block, subsampling the frames by a factor of 4.
A linear transformation to map the front-end output to the model's dimension is applied for all features.
Yet, depending on the feature extraction variant, the input dimension and therefore the size of the linear layer differs drastically.
The remaining \gls{am} consists of 12 Conformer blocks \cite{gulati2020conformer} with relative positional encodings \cite{shaw2018relativeposition}.
We use a hidden dimension of 512 and a feed-forward dimension of 2048, resulting in $\sim$ 77M parameters in the log Mel baseline.
For learning rate scheduling, we use a one cycle learning rate starting from $7\cdot10^{-6}$ and peaking at $7\cdot10^{-4}$ at the middle of training.
AdamW \cite{loshchilov2018decoupled} with a weight decay of 0.01 is deployed as the optimizer and the gradients' norm is clipped with a value of 1.0.
We train the model for 100 full epochs using speed perturbation with factors randomly sampled from \{0.9, 1.0, 1.1\}.
For recognition, we use the last checkpoint and decode our model with Flashlight \cite{kahn2022flashlight}.
Training is feasible on a single consumer GPU with \gigabyte{24} VRAM (e.g. Nvidia RTX 3090), meaning the entry barrier for reproduction is low.
Our work and code is publicly accessible.\footnote{\url{https://github.com/rwth-i6/returnn-experiments/tree/master/2025-2d-conv-features}}

\vspace{-0.5ex}
\subsection{SpecAugment}
\vspace{-0.5ex}
\label{sec:spec_aug}
A prevalent method used for regularization is SpecAugment \cite{park2019specaugment}, where randomly selected regions of a given input are masked in the time and feature dimensions during training.
In this work, we aim to unify the feature extraction front-end and remove influences from traditional methods.
In this context, it is desirable not to apply SpecAugment on an arbitrary intermediate layer.
Replacing the classical feature extraction 1:1 with a learnable front-end results in SpecAugment being applied between the feature extractor and the VGG-style convolutional block \cite{vieting2023itg}.
In contrast, generic positions are upfront before the feature extraction as in \cite{vieting2025features} or possibly between convolutional front-end and Conformer similar to the masking in wav2vec 2.0 \cite{baevski2020wav2vec2}.
In wav2vec 2.0, the masking position is inherently predetermined by the structure of the loss function in self-supervised pre-training and not altered for the supervised fine-tuning.

\begin{table}[htbp]

\centering
\begin{tabular}{|c|c|c|S[table-format=1.1]|S[table-format=1.1]|S[table-format=1.1]|S[table-format=1.1]|}
\hline
                   Feature &  \multirow{3}{*}{\shortstack{\#Params\\before\\Conformer}} & \multirow{3}{*}{\shortstack{Spec-\\Augment}} &                          \multicolumn{4}{c|}{{WER [\%]}} \\\cline{4-7}
                extraction &                                          \multirow{3}{*}{} &            \multirow{3}{*}{} & \multicolumn{2}{c|}{{dev}} & \multicolumn{2}{c|}{{test}} \\\cline{4-7}
              architecture &                                          \multirow{3}{*}{} &            \multirow{3}{*}{} & {clean} &          {other} & {clean} &           {other} \\\hline\hline
                   log Mel &                                            \phantom{0}1.4M &    \multirow{2}{*}{Features} &     2.4 &              5.2 &     2.7 &               5.6 \\\cline{1-2}\cline{4-7}
\multirow{2}{*}{\gls{scf}} &                                     \multirow{2}{*}{12.4M} &                              &     2.6 &              5.7 &     2.9 &               6.0 \\\cline{3-7}
                           &                                                            &  \multirow{4}{*}{\gls{stft}} &     2.6 &              5.6 &     3.0 &               6.0 \\\cline{1-2}\cline{4-7}
                   wav2vec &                                            \phantom{0}5.0M &                              &     2.6 &              5.9 &     2.9 &               6.3 \\\cline{1-2}\cline{4-7}
       \multirow{2}{*}{2D} &                                            \phantom{0}2.3M &                              &     2.5 &              5.5 &     2.9 &               5.9 \\\cline{2-2}\cline{4-7}
                           &                                            \phantom{0}0.3M &                              &     2.5 &              5.9 &     2.9 &               6.2 \\
\hline
\end{tabular}
\caption{
        Overview of different feature extraction methods.
        For the 2D features, the first layer is a randomly initialized filterbank with a kernel size of 256 and a stride of 10.
        The number of channels is 128 in the better-performing case and 8 in the parameter-efficient case.
        It is followed by 6 2D layers with a subsampling factor of 2 each.
        SpecAugment has been tuned for feature-level and \gls{stft}-domain separately, but is consistent across the different extractors.
}
\label{table:results_overview}
\vspace{-0.7em}
\end{table}

We deploy the first variant here because it has a few key advantages.
As explained in \cite{vieting2025features}, it avoids issues arising from the random order of filters in learnable features.
Furthermore, it is not possible for the model to spread information over multiple channels in order to bypass the masking.
Finally, the hyperparameters are independent of the feature dimension and do not need to be tuned when changing the feature extractor.

\section{Results}
First, we present \glspl{wer} for a comparison of the different described feature extraction methods in \Cref{table:results_overview}.
Unlike previous works \cite{vieting2023itg,vieting2025features}, there is a clear performance degradation when moving from log Mel features to \gls{scf}.
In contrast to \cite{vieting2025features}, we do not tune the audio perturbation and only applied speed perturbation with rather limited perturbation factors which might contribute to this observation.
Moving SpecAugment to the \gls{stft} domain results in the same performance.
Surprisingly, the wav2vec feature extractor is performing about 5\% relatively worse.
This is in contrast with \cite{vieting2021waveform,vieting2023itg} where it slightly outperformed \gls{scf}.
However, our generic unified 2D convolutional feature extractor performs on par with \gls{scf}.
This demonstrates that it is possible to largely reduce the influence of handcrafted features and rebuild a generic front-end from scratch.

Moreover, our proposed feature extractor is highly parameter-efficient as demonstrated in \Cref{table:results_overview}.
The reported number of parameters before the Conformer includes the feature extractors, a VGG-style module if applicable (for log Mel and \gls{scf}), and the linear layer mapping the output to the hidden dimension of the \gls{am}.
The number of parameters is significantly lower than for the other learnable front-ends even for the best-performing configuration.
With a parameter-efficient configuration, it can be reduced to only 0.3M parameters which is considerably smaller than even for log Mel.
Note that the parameters are distributed differently for the \gls{scf} and wav2vec front-ends, though.
For \gls{scf}, the bulk of these parameters is in the linear layer before the Conformer because the feature dimension is so large.
However, different methods to reduce the feature dimension clearly degraded the performance in preliminary experiments not presented here.
For the wav2vec feature extractor, most parameters are in the convolutional layers.

In addition, the learnable front-ends in \Cref{table:results_overview} increase the training time compared to the log Mel baseline.
The increases are 106\% for \gls{scf}, 25\% for the wav2vec feature extractor, and 57\% or 6\% for the proposed 2D convolutional front-end configurations.
The number of channels in the first layer thus allows trading off the \gls{wer} against the training time.
Further speedups are possible with higher subsampling in the first layer as studied below.

\begin{table}[htbp]

\centering
\glsunset{gt}

\begin{tabular}{|c|c|c|S[table-format=1.1]|S[table-format=1.1]|S[table-format=1.1]|S[table-format=1.1]|}
\hline
                                                   \multicolumn{3}{|c|}{First layer} &                          \multicolumn{4}{c|}{{WER [\%]}} \\\hline
      \multirow{2}{*}{Type} &                     Train- &     \multirow{2}{*}{Init} & \multicolumn{2}{c|}{{dev}} & \multicolumn{2}{c|}{{test}} \\\cline{4-7}
          \multirow{2}{*}{} &                       able &         \multirow{2}{*}{} & {clean} &          {other} & {clean} &           {other} \\\hline\hline
            \gls{stft} mag. &        \multirow{3}{*}{No} &        \multirow{2}{*}{-} &     2.5 &              5.5 &     2.9 &               5.9 \\\cline{1-1}\cline{4-7}
    \gls{stft} $[\Re,~\Im]$ &                            &                           &     2.4 &              5.6 &     2.9 &               6.0 \\\cline{1-1}\cline{3-7}
\multirow{3}{*}{Filterbank} &                            & \multirow{2}{*}{\gls{gt}} &     2.5 &              5.5 &     3.0 &               5.9 \\\cline{2-2}\cline{4-7}
                            &       \multirow{2}{*}{Yes} &                           &     2.4 &              5.4 &     2.9 &               5.9 \\\cline{3-7}
                            &                            &                    random &     2.5 &              5.6 &     2.9 &               5.9 \\
\hline
\end{tabular}
\caption{Comparison of different first layer types. All experiments use a window shift/stride that results in a subsampling factor of 10. The \gls{stft} uses a window size of 400 samples, the filterbanks use 80 filters of size 256. \Gls{stft} mag. and $[\Re, \Im]$ denote the magnitude or real and imaginary parts, respectively,  GT denotes the Gammatone filterbank. After the first layer, 6 2D layers with stride 2 are deployed.}
\label{table:results_first_layer}
\end{table}

\begin{figure}[htb]
	\centering
	\vspace{-0.5em}
	\includegraphics{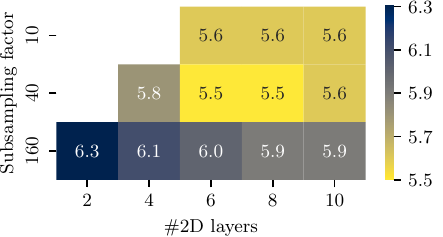}
	\caption{
		Interaction between subsampling factors in the first layer and number of subsequent 2D layers.
		All experiments use a randomly initialized learnable filterbank with 80 filters of length 256 here.
		There are 2/4/6 2D layers with a stride of 2 to obtain an overall subsampling factor of 640 in each case.
		Further 2D layers are added with a stride of 1.
		Results are \glspl{wer} on dev-other.
	}
	\label{fig:heatmap}
	\vspace{-1.0em}
\end{figure}

In the following, a number of ablation studies is presented in order to explore different aspects of the generic feature extraction.
First, \Cref{table:results_first_layer} compares different choices for the first layer, which needs to extend the 1D waveform with a feature dimension in addition to the time dimension. 
For this, we compare the \gls{stft} against different filterbank variations.
When using the \gls{stft}, the real and imaginary parts have the theoretical advantage of avoiding the information loss that is present in the case of the magnitude.
However, this does not translate to better \glspl{wer} in the experiments.
An improved fusion method of both parts could possibly improve the results, but the potential gain is likely to be small.

Moving to a filterbank as first layer, we can choose to train or freeze it during training and to initialize it with a Gammatone filterbank or randomly.
As shown in the last three rows of \Cref{table:results_first_layer}, the Gammatone initialization slightly helps on the dev sets.
However, the random initialization performs on par on the test sets, which is why we conclude that a filterbank of sufficient quality can be learned from scratch even in a generic setup.
To align with the goal of a monolithic neural network performing all steps needed for \gls{asr} starting from the waveform, we use a randomly initialized filterbank in the following experiments.

Next, we study the interaction between the amount of subsampling performed in the first layer and the numbers of subsequent 2D layers.
\Cref{fig:heatmap} shows the \gls{wer} for different combinations of subsampling factors and number of 2D layers.
In each row, the leftmost result uses the minimal number of layers with stride 2 required to achieve a final frame shift of \ms{40}.
Additional layers with a stride of 1 can be added in between to increase the total number of 2D layers.
The results show that lower subsampling factors and more layers are beneficial.
A plateau is reached after 6 2D layers and subsampling factors 10 and 40 perform similarly there.
Unlike for the learnable filterbank, the configuration did not have such a clear impact when using the \gls{stft} magnitude in experiments not presented here.

\begin{figure*}[htb]
    \centering
	\resizebox{\linewidth}{!}{
    \input{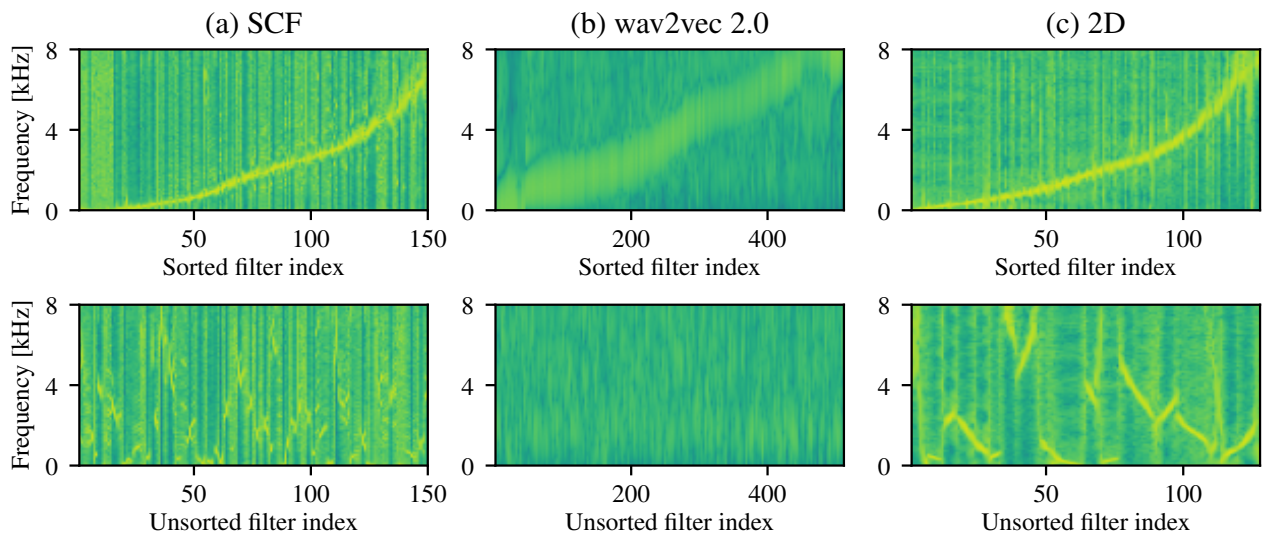}
	}
    \vspace{-2.2em}
    \caption{
		Frequency response of different feature extractors' filters in the first layer that operates on the waveform.
		The frequency responses are sorted by the peak frequency in the top row and in the order that the network learned them in the bottom row.
		For the proposed 2D convolutional front-end, we take the filters of the best-performing configuration in \Cref{table:results_overview}.
	}
    \label{fig:first_layer}
    \vspace{-1.2em}
\end{figure*}

\begin{figure}[htb]
	\centering
	\input{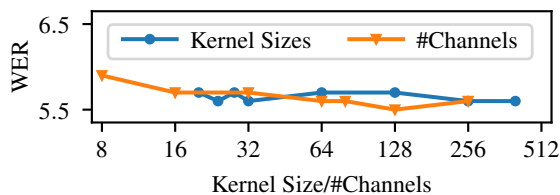}
	\caption{
		Comparison of different kernel sizes and numbers of channels in the first layer.
		All experiments use a randomly initialized filterbank with stride 10.
		For the blue curve, the number of channels is 80 while the kernel size is 256 for the orange curve.
		After the first layer, 6 2D layers with stride 2 are deployed.
		Results are \glspl{wer} on dev-other.
		The model with kernel size 16 did not converge.
	}
	\label{fig:conv_kernelsize_numchannels}
\end{figure}

Our last study deals with the kernel size and the number of channels in the filterbank, investigating how much we can reduce the number of parameters without significantly degrading the performance.
The results are shown in \Cref{fig:conv_kernelsize_numchannels}.
Reducing the kernel size from 400 --which is the \gls{stft} window size in our experiments-- to 256 --which is the kernel size in the first \gls{scf} layer-- and further below does not have a significant impact on the performance.
Only dropping to a size of 16 introduces a severe impairment.
Reducing the number of channels degrades the \gls{wer} only mildly.
Interestingly, this observation is fairly consistent with long-established beliefs in the research field where for Gammatone filters, the minimum length is specified as 32 and the minimum number of channels as 8 \cite{patterson1987gammatone}.

\subsection{Analysis of Learned Filters}
One way to compare the different filters learned in the first layer is by plotting their frequency responses.
The plots can be seen in \Cref{fig:first_layer}.
Given that the order of the filters learned by the neural network is arbitrary in general, we sort the filters by the peak frequency followed by the upper and lower \SI{3}{\decibel} cutoff frequencies as second and third sorting criteria in the top row.
The frequency responses for \gls{scf} and wav2vec feature extractor are consistent with the findings in \cite{vieting2023itg}, although it appears that there are fewer unused \gls{scf} filters here compared to \cite{vieting2023itg}.
Similar to \gls{scf}, the first layer filterbank in our proposed generic feature extractor using 2D convolutions also learns a set of bandpass filters whose characteristics are similar to the Gammatone filterbank.

However, a notable difference can be observed when inspecting the unsorted frequency responses in the bottom row of \Cref{fig:first_layer}.
While the filter order appears to be random for \gls{scf} and wav2vec feature extractor, the filters learned in the first layer of our proposed feature extractor show clear groups of filters with adjacent ascending or descending center frequencies.
This can likely be explained by the nature of the subsequent 2D convolutions with small kernel sizes.
When operating with a $3\times3$ kernel, it can be advantageous for the 2D layers to have related information such as similar frequency bands in neighboring feature channels since only then those channels are seen together by the kernel.

\section{Limitations and Future Work}
While this paper demonstrates the feasibility of a unified generic front-end to extract features for \gls{asr}, the performance of the learnable front-ends falls short of log Mel features.
This could be at least partially caused by a lack of audio perturbation.
We only use speed perturbation with factors sampled from the fixed set \{0.9, 1.0, 1.1\} which might not have sufficiently strong regularization effect.
In contrast, \cite{vieting2025features} showed improvements when using tempo perturbation with stronger perturbation factors sampled from a continuous distribution.
However, tuning the audio perturbation is out of the scope of this work.
Another reason might be the low tuning effort for SpecAugment compared to the log Mel baseline while various other aspects of the pipeline such as the learning rate schedule are not tuned at all.
Nevertheless, the proposed generic architecture and the insights gained in this work open up many possibilities for further studies on features trained from scratch, possibly also for other tasks besides \gls{asr}.

\vspace{-1.0em}
\section{Conclusions}
In this work, we present a unified generic feature extraction architecture for \gls{asr} based on 2D convolutions.
The design aims to minimize the influence of traditional feature extraction methods and to unify previously existing architectural inconsistencies in the convolutional front-end.
The results using a \gls{ctc} \gls{asr} system on LibriSpeech prove that the proposed architecture is feasible and the remaining difference to log Mel features is not due to its missing structure.
This is evidenced by the competitive performance with existing learnable feature extractors.
The ablations show that both the \gls{stft} and a learnable filterbank can be used for time frequency decomposition in the first layer with similar performance.
Furthermore, the front-end is highly parameter-efficient especially for a filterbank with a low number of channels in the first layer, allowing to trade off training speed vs. \gls{wer}.
Finally, our analyses demonstrate that even with the generic design, the front-end exhibits behaviors that align with well-established observations in the field.

\vspace{-1.0em}
\section{Acknowledgements}
This work was partially supported by NeuroSys, which as part of the initiative “Clusters4Future” is funded by the Federal Ministry of Education and Research BMBF (03ZU2106DD), and by the project RESCALE within the program \textit{AI Lighthouse Projects for the Environment, Climate, Nature and Resources} funded by the Federal Ministry for the Environment, Nature Conservation, Nuclear Safety and Consumer Protection (BMUV), funding IDs: 67KI32006A.

\small
\bibliographystyle{IEEEtran}
\bibliography{references}

\end{document}